\documentclass[useAMS,usenatbib]{mn2e}

\usepackage{graphicx}

\title[Supernova dust at $z \sim 1$]{
Supernova dust for the extinction law in a young infrared galaxy at $z \sim 1$}
\author[K. Kawara et al.]{K. Kawara$^{1}$\thanks{E-mail:
kkawara@ioa.s.u-tokyo.ac.jp}, H. Hirashita,$^{2}$ T. Nozawa$^{3}$, T. Kozasa$^{4}$, 
\newauthor
S. Oyabu$^{5}$, Y. Matsuoka$^{5}$, T. Shimizu$^{1}$, H. Sameshima$^{1}$, and N. Ienaka$^{1}$\\
$^{1}$Institute of Astronomy, the University of Tokyo, Osawa 2-21-1, Mitaka, 
Tokyo 181-0015, Japan\\
$^{2}$Institute of Astronomy and Astrophysics, Academia Sinica, 
PO Box 23-141, Taipei 10617, Taiwan\\
$^{3}$Institute for the Physics and Mathematics of the Universe,
the University of Tokyo,
 5-1-5 Kashiwanoha, Kashiwa, 277-8568, Japan\\
$^{4}$Department of Cosmosciences, Hokkaido University, Sapporo 060-0810, 
Japan\\
$^{5}$Graduate School of Science, Nagoya University, Furo-cho, Chikusa-ku, 
      Nagoya 464-8602, Japan\\
}
\begin{document}

\date{Accepted in MNRAS 2010 November 2; in original form 2010 March 1}
%%\date{Accepted 2009 March 15. Received 2009 June 14; in original form 1988 
%%Octobe 11}

\pagerange{\pageref{firstpage}--\pageref{lastpage}} \pubyear{2002}

\maketitle

\label{firstpage}

\begin{abstract}
We apply the supernova(SN) extinction curves to reproduce the observed properties of SST J1604+4304 
which is a young infrared (IR) galaxy at $z \sim 1$. 
The SN extinction curves used in this work were obtained 
from models of unmixed ejecta of type II supernovae(SNe II) for the Salpeter initial mass function (IMF) 
with a mass range from 8 to 30 $M_{\sun}$ or 8 to 40 $M_{\sun}$. 

The effect of dust distributions on the attenuation of starlight is investigated by performing 
the $\chi^2$ fitting method against various dust distributions. These are the commonly used uniform dust 
screen, the clumpy dust screen, and the internal dust geometry. We add to these geometries
three scattering properties, namely, no-scattering, isotropic scattering, 
and forward-only scattering. Judging from the  $\chi^2$ values, we find that
the uniform screen models with any scattering property provide good approximations to the real 
dust geometry. Internal dust is inefficient to attenuate starlight and thus cannot be the dominant 
source of the extinction.

We show that the SN extinction curves reproduce the data of SST J1604+4304 comparable to or 
better than the Calzetti extinction curve. The Milky Way extinction curve is not in satisfactory 
agreement with the data unless several dusty clumps are in the line of sight. 
This trend may be explained by 
the abundance of SN-origin dust in these galaxies; SN dust is the most abundant in the 
young IR galaxy at $z \sim 1$, abundant in local starbursts, and less abundant in the Galaxy.
If dust in SST J1604+4304 is dominated by SN dust, the dust production rate is 
$\sim 0.1$ M$_{\sun}$/SN. 

\end{abstract}


\section{Introduction}

Dust is a minor constituent in the Universe, but it plays crucial roles for evolution of 
various objects; for example, it cools gas clouds by radiating far-infrared (far-IR) emission, 
and acts as a catalyzer to form molecules. 
Corrections for dust extinction have significant cosmological consequences 
for determining the Hubble constant, the cosmological model, and the epoch of galaxy formation.

It is generally accepted that the atmospheres of asymptotic giant branch(AGB) stars are one of 
the main sources of interstellar dust in the Milky Way (MW). However, AGB stars are too old to 
account for the presence of dust at $z > 6$ (but see \citealt{valiante09}), 
and supernovae (SNe) have been recognized as a 
candidate of the major source of dust in the early stage of galaxy evolution.

Stars with masses 8--40 $M_\odot$ end up their life as type II supernovae(SNe II)\footnote{
In this paper, we simply use the term
SNe II for core-collapsed SNe whose progenitors are massive stars.
We also call the dust produced by SNe II ``SN dust'',
and the extinction curve synthesized for the SN dust``SN extinction curve''.} in $\le$
40 Myr since their formation \citep{heger03}. The formation of 
dust in SN ejecta was
first witnessed in SN 1987A, motivating \citet{kozasa89} and \citet{kozasa91} 
to predict the condensed dust composition and size.
Systematic studies of dust formation in SN ejecta have been 
proceeded to explore the role of dust in the formation and evolution of 
stars and galaxies in the early universe (\citealt{todini01}; \citealt{nozawa03}; 
\citealt{schneider04}): 
the composition of synthesized dust reflects the elemental composition 
in the formation site which is affected by the type and degree of mixing 
and the degree of formation efficiency of CO molecule, and the size of
dust is controlled by the time evolution of gas density and temperature
depending on the type of SNe (\citealt{nozawa08};\citealt{nozawa10}). 

Apart from the differences in the SN models and 
the underlying assumptions employed for dust formation calculations, the
predicted dust mass formed in the ejecta reaches 0.1 to 1$M_\odot$ per
SN II, which is sufficient enough to explain a vast amount of dust of 
$ > 10^8M_{\odot}$  discovered in host galaxies of 
high-redshifted quasars (QSOs) at $z>5$(eg. \citealt{morgan03me}; \citealt{maiolino06}; 
\citealt{dwek07}). However, the mass and composition of dust
formed in the ejecta have been still controversial; the observations of 
nearby SNe have claimed dust mass formed in the ejecta to be less than $10^{-3}
M_\odot$ (see \citealt{kozasa09} for a review), while 
the infrared and submillimeter observations of Galactic supernova remnants (SNRs)
have suggested the presence of dust reaching 0.02 to 1 $M_\odot$ (\citealt{rho08}; 
\citealt{dunne03}; \citealt{dunne09}; \citealt{gomez09}; \citealt{temim10}; \citealt{nozawa10}). 
Although no information 
has been available on the composition of dust formed in the ejecta except for SN 2004et 
\citep{kotak09}, the spectral mapping observation covering the
entire of Cas A SNR with the {\it Spitzer Space Telescope} have revealed
the presence of a variety of dust species whose composition is closely
associated with the gas emission line characteristic of
nucleosynthetic layer in the progenitor \citep{rho08}.

The amount, composition and size of dust
supplied from SNe to the interstellar medium (ISM) are not always the same as those condensed in
the ejecta due to the processing in a hot plasma produced by the
reverse and forward shocks during the journey to the ISM (\citealt{bianchi07}; 
\citealt{nozawa07}) as well as in the hot/warm
region of ISM (\citealt{jones94}; \citealt{jones96}; \citealt{nozawa06}; 
\citealt{hirashita10}).
Nevertheless \citet{maiolino04} found that the extinction curve of SDSS J1048+4647, a
broad absorption line (BAL) QSO at $z$=6.2 is different from those of
nearby galaxies, and is in excellent agreement with the extinction curve
obtained from SN dust models by \citet{todini01}
considering the uniformly--mixed elemental composition in the ejecta.
Synthesizing dust extinction curves for various SN dust models by
\citet{nozawa03}, \citet{hirashita05} have shown that the
extinction curve observed in
this BAL QSO can be reasonably reproduced by the dust produced in SNe
II without any mixing in the ejecta.  Also the extinction curve derived
from the analysis of the
afterglow of $\gamma$--ray burst GRB 050904 at $z=6.3$ have suggested
that SNe are the dominant source of dust in the host galaxy \citep{stratta07}, 
although \citet{zafar10} have
claimed no evidence of dust extinction in the afterglow.
However, the observed extinction curve in the local environments may provide less constraint
on the SN dust model as well as the role of SNe on the evolution
of galaxies.

Young galaxies, in which no AGB stars evolved off the main sequence, are obvious targets to study 
the properties of SN dust, for example, extinction curves that determine the relation between changes 
in color and total absorption. However, it is generally difficult to infer the 
amount of the dust obscuration, because different geometries of the dust spacial distribution 
result in different observed colors for the same amount of dust and the same intrinsic spectrum 
of the underlying stellar population(e.g., \citealt{giavalisco02}). In fact, some studies of HI 
recombination lines in starburst and dusty Seyfert galaxies suggest that the extinction observed 
in the near-IR is generally greater than that observed by using the optical lines 
(\citealt{kawara89}; \citealt{puxley94}). 

\citet{calzetti94} has shown that the extinction laws in the Milky Way and Large Magellanic Cloud 
cannot account for the ultraviolet (UV) to optical spectra of UV-bright starburst 
galaxies in the local Universe in a satisfactory way. They empirically derived the extinction law 
for these galaxies. Their extinction law is characterized by an overall slope which is more gray or 
flatter than the MW extinction law's slope and by the absence of the 2175 \AA\ dust feature. 
In starburst galaxies, numerous SN II explosions occur and SN-condensed dust would be injected into the 
ISM. Thus, newly formed SN dust would produce a great impact on the extinction law
in local starburts.

In this paper, we apply synthetic extinction curves on the unmixed ejecta models of 
SNe II by \citet{hirashita05} to the broad-band SED of a young ultra luminous IR galaxy (ULIRG) 
SST J1604+4304 at $z \sim 1$. 
Comparing with the MW and \citet{calzetti01} extinction laws, we show that the SN 
extinction curves fit better to the observations. 
 We adopt $H_0$ = 70 km s$^{-1}$ Mpc$^{-1}$, 
$\Omega_m = 0.3$, and $\Omega_{\Lambda} = 1 - \Omega_m$ throughout this paper. 

\begin{figure}
%%\epsscale{0.8}
%%\includegraphics[angle=0,scale=1.0]{k3_ulirgs.ps}
\includegraphics[width=84mm]{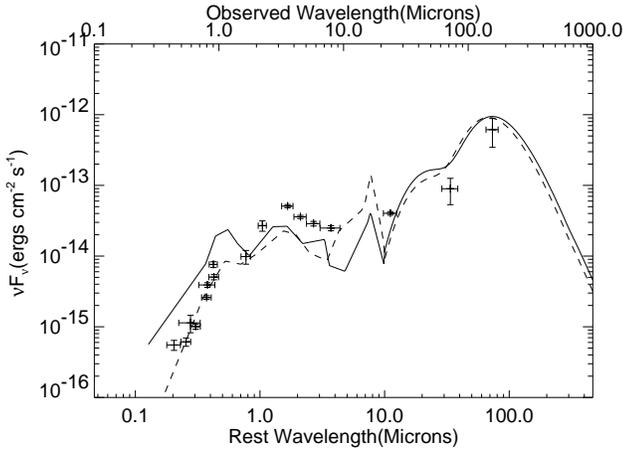}
\caption{UV to far-IR SED for SST J1604+4304 compared with those of 
starburst-dominated ULIRGs, Arp 220({\it solid} local ULIRG), 
MIPS J142824.0+352619({\it dash} $z = 1.3$ hyperluminous IR galaxy). 
Arp 220 data come from \citet{carico90}, 
\citet{rigopoulou96}, and \citet{klaas97}, and MIPS J142824.0+352619 from \citet
{desai06} and \citet{borys06}.
}
\label{f_ulirgs}
\end{figure}

\section[]{Data}

Synthetic spectra of stellar populations reddened by dust are modelled to reproduce the data of 
SST J1604+4304. This is a young IR galaxy at $z = 1.135$ with the characteristic of ULIRGs. 
The multi-wavelength data from the X-ray to radio are compiled in \citet{kawara10}. 
There are no supportive evidence for AGN activities in this galaxy, although the data are not 
deep enough to rule out the presence of an AGN.

Figure \ref{f_ulirgs} illustrates the broad-band SED from the UV to far-IR\footnote{The data 
are taken from Table 1 in \citet{kawara10} and the correction for 
the gravitational amplification is applied.}  
along with representative SED templates of starburst-dominated ULIRGs, namely,
Arp220(local ULIRG) and MIPS J142824.0+352619 (hyper-luminous IR galaxy at $z = 1.3$). 
The broad-band SED and stellar absorption lines, such as CaII and HI lines, as well as 
the [OII]$\lambda$3727 line suggest that young stellar populations dominate the optical 
luminosity in this galaxy. The CaII index, defined by \citet{kawara10}, is D(CaII) = 0.03 $\pm$ 
0.11 and the equivalent [OII] width is EW([OII]) = 31.1 $\pm$ 4.3 \AA\ in the rest-frame which 
corresponds to the observed [OII] line 
flux\footnote{The [OII] line flux published in \citet{kawara10} is incorrect.} 
of $1.9 \pm 0.3 \times 10^{-17}$ ergs cm$^{-2}$ s$^{-1}$.

\citet{kawara10} estimated the reddening to be $E(B-V) \sim 0.8$ 
when the Calzetti extinction law is applied. 
The 8-1000\micron\ bolometric luminosity inferred from the three $Spitzer$ MIPS bands at 
the 24, 70, and 160\micron\ is $L_{ir} = 1.78 \pm 0.63 \times 10^{12} L_{\sun}$, corresponding to
the flux $F_{ir} = 1.00 \pm 0.35 \times 10^{-12}$ ergs cm$^{-2}$s$^{-1}$.
These values are intrinsic to the galaxy after correcting for a gravitational amplification of 1.17
by the foreground galaxy cluster CL 1604+4304 at $z = 0.9$. 
The bolometric luminosity agrees with the extinction-corrected, stellar luminosity, 
suggesting SST J1604+4304 can be characterized by models of pure star formation with no AGN 
emission. The dust mass inferred from the far-IR data is $\sim 2 \times 10^8$ M$_{\sun}$, 
sufficient to form a shell surrounding the galaxy and produce the observed extinction.

The SED data at 13 photometric bands from $B$ to 5.8\micron\ 
are used for the following analysis; the longest $Spitzer$ IRAC 8.0\micron\ band is not used, 
because 
the dust emission may be greater than the starlight in this band corresponding to 3.7\micron\ in 
the rest-frame. 

\section{Extinction}
\subsection{Geometry of dust distribution}

The attenuation produced by dust in front of an extended source does not only depend on the 
properties 
of dust grains, such as cross sections, albedo, and amount, but also depends on the geometry of 
spacial distribution of dust. It is known that the apparent optical depth is lower than the optical 
depth averaged over the source area when starlight is attenuated by a thick, clumpy dust layer or 
dust is internal to the extended source. We consider the following geometries of dust distribution. 
For full details, the reader may refer to \citet{code73},\citet{natta84}, and \citet{calzetti94}. 
 
{\it Internal dust} -- In the case that absorbers(dust grains) and emitters(stars) are uniformly 
distributed, the apparent extinction is given by:
\begin{equation}
f_{\nu} = \epsilon_{in}f^0_{\nu} \textrm{\ \ \ \ \ \ \ \ \ with\ \ } 
        \epsilon_{in} = e^{-\tau_{ap}} \\
\end{equation}

\begin{equation}
\tau_{ap} = -\ln[(1 - e^{-\tau_{eff}})/\tau_{eff}] \\
\end{equation}

\begin{equation}
\tau_{eff} =  (1-\gamma)\tau_{ext}        \textrm{\ \ \ \ \ \ \ \ \ for $g = 1$}
\end{equation}
\begin{equation}
\tau_{eff} =  (1-\gamma)^{0.75}\tau_{ext}  \textrm{\ \ \ \ \ for $g = 0$}
\end{equation}    
where $\tau_{ap}$ is the apparent optical depth of the internal dust,
$\tau_{eff}$ the effective absorption optical depth, 
$\tau_{ext}$ the sum of the optical depth for absorption and scattering, $\gamma$
the albedo, $f_{\nu}^0$ is the unobscured, intrinsic flux density , and 
$f_{\nu}$ is the reddened, observed flux density. $g \equiv <cos(\theta)>$ is the scattering 
asymmetry factor, where $g$ = 1 for the forward-only scattering and $g$ = 0 for 
the isotropic scattering. Sometimes $\tau_{eff} =  (1-\gamma)^{0.5}\tau_{ext}$ is used for the 
isotropic scattering (i.e., \citealt{natta84}; \citealt{calzetti94}). 
However, $\tau_{eff} =  (1-\gamma)^{0.75}\tau_{ext}$ approximates better to the solution by the 
two-stream approximation for the geometry with a uniform distribution of the absorbers 
and the emitters. For the full details, refer to Figure \ref{f_appendix} in Appendix A.  

{\it Dust screen} -- In this geometry, the absorbers are physically separated from and located 
in a screen in front of the emitters. We adopt an analytic expression derived by \citet{natta84} 
for a system of clumps distributed according to Poisson statistics.
\begin{equation}
f_{\nu} = \epsilon_{sc}f^0_{\nu} \textrm{\ \ \ \ \ \ \ \ \ with\ \ } 
        \epsilon_{sc} = e^{-\tau_{ap}}\\
\end{equation}
\begin{equation}
\tau_{ap} =  N(1-e^{-\tau_{eff}/N})   \\
\end{equation}
where $N$ is the average number of clumps along the line of sight and all clumps are assumed to 
have one and the same effective optical depth $\tau_{eff}/N$. For $N \rightarrow \infty$, 
$\tau_{ap} \rightarrow  \tau_{eff}$, corresponding to the uniform dust screen. For the dust 
screen which is physically distant from the system of stars, the effect of the dust is to remove 
photons from the line of sight through absorption and scattering. 
On the other hand, for the dust screen which is located close to the system of stars, the scattered 
photons can get in the line of sight. Thus, $\tau_{eff}$ is given as:

\begin{equation}
\tau_{eff} =  \tau_{ext}        \textrm{\ \ \ \ \ \ \ \ \ \ \ \ \ \ \ \ \ for no scattering}
\end{equation}
\begin{equation}
\tau_{eff} =  (1-\gamma)\tau_{ext}        \textrm{\ \ \ \ \ \ \ \ \ for scattering with $g = 1$}
\end{equation}
\begin{equation}
\tau_{eff} =  (1-\gamma)^{0.5}\tau_{ext}  \textrm{\ \ \ \ \ for scattering with $g = 0$}
\end{equation}    
Again, $g = 1$ is for the forward-only scattering, and  $g = 0$ for the isotropic scattering.

{\it Composite geometry} -- It is quite likely that dust grains distribute within the system of 
stars as internal dust and in the screen in front of the system. We approximate this case as:
\begin{equation}
f_{\nu} = \epsilon_{sc}\epsilon_{in}f^0_{\nu} \\
\end{equation}

\begin{figure}
%%\epsscale{0.8}
\includegraphics[width=84mm]{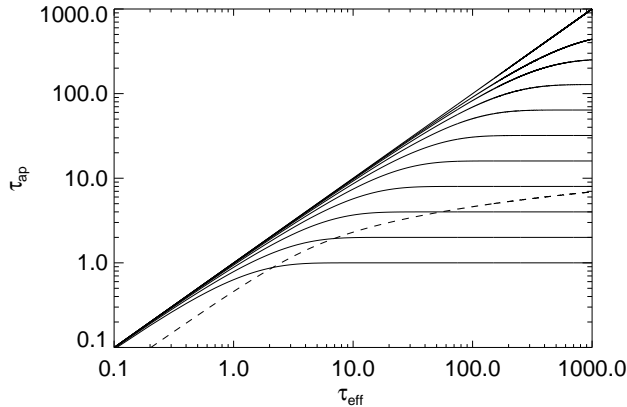}
%%\includegraphics[angle=0,scale=1.0]{k3_chart.ps}
%%\plotone{c1604_image_zoom.ps}
\caption{The apparent optical depth $\tau_{ap}$ is shown as a function of the effective optical depth 
$\tau_{eff}$. The solid lines represent dust screen geometries for $N$ = [1,2,4,8,16,32,64,128,256,
512,$\infty$] from bottom to top. The dashed line represents an internal dust geometry.
} 
\label{f_clumps}
\end{figure}

Figure \ref{f_clumps} shows the apparent optical depth $\tau_{ap}$ as a function of $\tau_{eff}$.
It is apparent that the growth of $\tau_{ap}$ turns over at a smaller value of  $\tau_{eff}$ 
in the dust screen with a smaller value of $N$. As shown in this figure, the extinction by dust 
in the screen with a small $N$ or by internal dust strongly depends 
on the geometry and tends to produce a gray extinction, and dust in a uniform screen produces the 
most effective extinction.

\subsection{Extinction curves}

\begin{figure}
%%\epsscale{0.8}
\includegraphics[width=84mm]{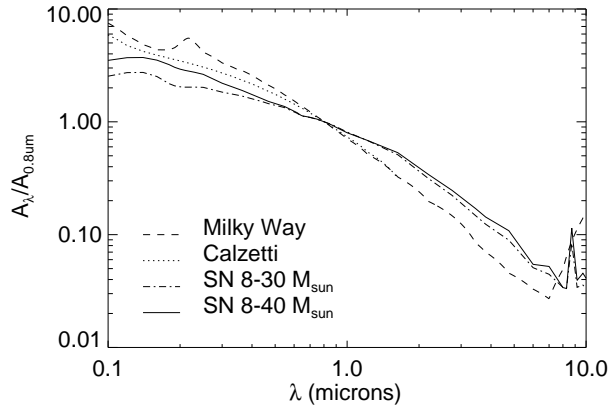}
%%\includegraphics[angle=0,scale=1.0]{k3_chart.ps}
%%\plotone{c1604_image_zoom.ps}
\caption{Synthetic SN extinction curve compared with those for the Milky way \citep{draine03} and 
local starbursts \citep{calzetti01}. 
The curves are normalized to the extinction (absorption + scattering) in magnitude at 0.8
\micron. Note that a spike on the SN curve is produced by crystal SiO$_2$ at 8.7 \micron.
} 
\label{f_curves}
\end{figure}

We examine two extinction curves synthesized in ejecta of SNe II, the empirical extinction curve 
derived by \citet{calzetti01}, and the MW extinction curve 
given by \citet{draine03}. Figure \ref{f_curves} plots these extinction curves as a function of 
wavelengths, showing that the SN extinction curves are more gray or flatter 
than the Calzetti and MW extinction curves from 0.1 - 1\micron.

The MW extinction curve exhibits a pronounced feature at 2175 \AA. This feature is absent in the 
typical extinction curves of the Small Magellanic Clouds (SMC). The dust in local starbursts and 
Lyman break galaxies exhibits the properties of the SMC-like dust properties, while \citet{noll09}
claimed that about 30\% of the spectra of actively star-forming, intermediate-mass galaxies at 
$ 1 < z < 2.5$ exhibit broad absorption features at $\sim $2175 \AA. 
To check the effect of the feature on the extinction law in SST J1604+4303, in addition to the MW 
extinction curve by \citet{draine03}, we will examine the pseudo MW extinction curve which 
is made by removing the 2175 \AA\ feature from the MW extinction curve.

The \citet{calzetti01} extinction curve was derived by analyzing local starbursts 
and blue compact dwarf galaxies \citep{calzetti94}, 
and is widely used to, for example, derive photometric redshifts of high-redshift galaxies.
\citet{calzetti01} gives the extinction curve in a narrow range from 0.12 to 2.2 \micron. 
To analyze multi-wavelength data of 
high-redshift objects from the UV to IR such as given in Figure \ref{f_ulirgs}, 
the extinction curve should be extended to 
longward of 2.2 \micron.  The functional form given by \citet{calzetti01} 
cannot be used to extend the extinction curve, because the extinction value sharply drop 
beyond 2.2 \micron. The Calzetti
extinction curve is almost identical to the \citet{draine03}'s MW curve from 0.6 to 1.6 \micron.  
We thus extend the Calzetti curve to longer wavelengths
by adopting the MW extinction curve for wavelengths $ \ge$ 1.6 \micron.  
In addition, because  \citet{calzetti01} does not provide the albedo, we adopt the albedo 
given in \citet{draine03} with removing the dip in the albedo which 
is attributed to the familiar 2175 \AA\ feature. This curve is called the Calzetti extinction 
curve in this paper. It is quite likely that the Calzetti extinction curve reflects the dust 
composition from the two origins, namely, dust condensed in expanding atmospheres of AGB stars and 
ejecta of SNe II, because in local starburt galaxies low-mass stars have already evolved off the 
main sequence and reached to the dust forming stages. Because the Calzetti curve was empirically 
obtained by assuming a geometry of uniform foreground screens (i.e., $f_{\nu} = e^{-\tau}f^0_{\nu}$),
we apply this curve in this form only. 

The two SN extinction curves are compared with the MW and Calzetti extinction curves.
The SN extinction curves were synthesized in unmixed ejecta of SNe II by \citet{hirashita05}, 
based on the dust production models by \citet{nozawa03}. 
\citet{nozawa03} have investigated the dust condensation in SNe II with various progenitor masses,
taking into account the time evolution of gas temperatures in the ejecta 
by solving the radiative transfer equation with energy deposition of 
radioactive elements.  The two extreme cases are considered for the elemental composition in 
the He-core: the unmixed case and the uniformly mixed case.
The formation of all possible condensations is calculated simultaneously by applying a theory 
of non-steady state nucleation and grain growth under the assumption of complete formation of 
CO and SiO molecules.  Various dust species condense according to
the elemental composition at the formation site in the unmixed ejecta;
Carbon in the inner C--rich He layer, Al$_2$O$_3$,
Mg-silicates (Mg$_2$SiO$_4$ and
MgSiO$_3$) in the oxygen--rich layer, MgO in the outer
O--rich layer, SiO$_2$ in the inner O--rich layer,
Si and FeS in the Si--S layer,
and FeS and Fe in the inner most Fe--S layer.
In the mixed ejecta, only oxide grains such as Al$_2$O$_3$, Mg--silcaites,
SiO$_2$ and Fe$_3$O$_4$
condense because the elemental composition assuming complete formation
of CO is oxygen--rich in the entire He--core.
In the unmixed cases, C and Si dust species dominate the dust mass in the low--mass 
progenitor \citep{nozawa03}, 
and Mg$_2$SiO$_4$ is getting dominant with
increasing progenitor mass, while in the mixed ejecta SiO$_2$ and
Mg$_2$SiO$_4$ dust dominate the dust mass.

Combining the optical constants in the literature with the grain species and size distribution of 
SN-condensed dust derived by \citet{nozawa03}, and using the Mie theory, 
\citet{hirashita05} have calculated absorption and scattering cross 
sections of homogeneous spherical grains with various sizes condensed in mixed and unmixed ejecta.
It should be pointed out here
that in the unmixed ejecta C and Si dust grains with radius $a \ge 0.1 \mu$m are
significantly abundant in the low mass progenitors than in the
high mass progenitors. This 
results in the different behavior of the extinction curves:
the extinction curve in UV region tends to be
flatter in the unmixed cases than in the mixed cases \citep{hirashita05}.
The details for the adopted quantities are described in \citet{hirashita05}.
The extinction curves in unmixed ejecta of SNe II reasonably reproduce 
the extinction law of the $z = 6.2$ BAL QSO found by \citet{maiolino04}.
In this paper, we use the two extinction curves of unmixed SNe II for 
the \citet{salpeter55} Initial Mass Function (IMF); one has a mass range of progenitors 
from 8 to 30 $M_{\sun}$, and the other from 8 to 40 $M_{\sun}$. These were obtained by 
interpolating or extrapolating extinction curves that \citet{hirashita05} 
calculated for unmixed SNe II with progenitor masses of 13, 20, 25, \& 30 $M_{\sun}$\footnote{
In \citet{hirashita05}, the slope of extinction curve for 25 M$_{\sun}$ and
30 M$_{\sun}$ progenitors was overestimated because of the error in the calculation processes.
We have replaced the results with correct ones. Nevertheless, the conclusions in \citet{hirashita05}
are not altered by this change.}.

\section[]{Synthetic spectra of stellar populations}

The evolutionary synthesis codes by \citet{bruzual03} are used to generate synthetic spectra 
of evolutionary stellar population models. We adopt the \citet{salpeter55} initial mass function (IMF)
with the lower and upper mass cutoffs of 0.1$M_{\sun}$ and 100$M_{\sun}$, respectively.
It is noted that the weight of massive stars is greater in the Chabrier IMF than in the Salpeter IMF,
resulting in a mass to luminosity ratio roughly 1.6 times heavier in the Salpeter IMF.
The star formation histories are exponentially declining models with
the star formation rate (SFR) $\propto$ $t_{sfr}^{-1}$exp(-$t/t_{sfr}$),
where $t_{sfr}$ is the e-folding timescale and $t$ is the age, i.e.,the time after the onset of 
the first star formation. Although \citet{kawara10} used the instantaneous burst models, such models 
are not appropriate to reproduce simultaneously the strong [OII]$\lambda$ 3727 emission line from 
HII regions ionized by young, hot stars and the observed SED.
The synthetic spectra are generated in a parameter range of 
$t = 3\times 10^5 - 8\times 10^9$ yr in a step of 0.2 dex and 
$t_{sfr} = 10^6 - 1.6 \times 10^{10}$ yr in a step of 0.3 dex. For the shortest $t_{sfr}$,
the star formation history is near an instantaneous burst, while for the longest $t_{sfr}$,
the star formation rate is almost constant over the cosmic time. 
The metallicity of the synthetic spectra has a range of $Z$ = 0.005 - 2.5 $Z_{\odot}$ 
(i.e., six points at 0.005, 0.02, 0.2, 0.4, 1.0, \& 2.5 $Z_{\sun}$).

The extinction models discussed in section 3 are then applied to the synthetic spectra generated in the 
above procedure, yielding template spectra which are used for the subsequent $\chi^2$-fitting. 
In the initial run, the extinction measured at 0.3 \micron\ 
has a range of $\tau_{ext}(0.3\micron) = 0 -10$ in a step of 0.1. In case 
that the initial range is too narrow to have the minimum $\chi^2$ for a certain model, the range and 
the step are widen by a factor of three to search for the minimum $\chi^2$, and so on. 

\section[]{$\chi^2$ template-fitting method}

We search for the best set of parameters for SST J1604+4304 by using a $\chi^2$ template-fitting 
method as defined as below:

\begin{equation}
\chi^2 = \chi_{SED}^2 + \chi^2_{[OII]} + \chi^2_{CaII} + \chi^2_{fir}
\end{equation}
where the degree of freedom is $\nu$ = 12. In what follows, we define the terms in equation (11). 

\begin{equation}
\chi_{SED}^2 = \sum_{filters}(F_{obs} - bF_{template})^2/\sigma_{SED}^2
\end{equation}
where $b$ is the parameter determined by fitting the template flux density $F_{template}$ to 
the observed $F_{obs}$. 
$F_{obs}$ are the photometric data measured in 13 broad band filters from $B$ to $5.8\micron$.
The $Spitzer$ IRAC 8.0 $\micron$ is not used, because dust emission may be significant in this 
band corresponding to 3.7 $\micron$ in the rest-frame. It is noted that 5\% of the flux was added 
in the quadratic form to the photometric error $\sigma_{ph}$, i.e., 
$\sigma_{SED}^2 = \sigma_{ph}^2 + (0.05F_{obs})^2$. This is necessary to avoid too small and 
non-realistic photometric error.  

The strong [OII]$\lambda$3727 line requires the presence of OB stars. We thus include $\chi_{[OII]}^2$ 
in the $\chi^2$ template-fitting method. $\chi^2_{[OII]}$ is defined as:
\begin{equation}
\chi_{[OII]}^2 = (EW^{[OII]}_{obs} - EW^{[OII]}_{template})^2/\sigma_{[OII]}^2
\end{equation}  
where $EW^{[OII]}_{obs} = 31.1 \pm 4.3$ \AA. 
To derive $EW^{[OII]}_{template}$, we first obtain the H$_{\alpha}$ luminosity by counting 
ionizing photons in the synthesized spectra. We then adopt the [OII] and H$_{\alpha}$ calibration 
performed by \citet{kewley04}. For the metallicity scaling of [OII]/H$_{\alpha}$, we use their theoretical 
curves for $q = 3 \times 10^7$ cm s$^{-1}$ with an additional condition 
that [OII]/H$_{\alpha}$ should not be less than 0.1. This additional condition is applied to avoid a 
negative or unrealistic value of [OII]/H$_{\alpha}$ outside of the calibrated range of metallicity. 
The calibrated [OII]/H$_{\alpha}$ has a peak value of 1.6 at $Z = 0.4 Z{\sun}$ and drops to a value 
of 0.1 at $0.14 Z{\sun}$ or $Z = 2.44 Z{\sun}$. Because \citet{kewley04} estimated the uncertainty 
to be $\sim$ 30\%, we add this uncertainity to the observational error as $\sigma_{SED}^2$ = 
$\sigma_{[OII] SED}^2$ + $(0.3EW^{[OII]}_{template})^2$.

As discussed in \citet{kawara10}, the D(CaII) index, which is the ratio of the CaII to HI absorption lines 
in equivalent width, is useful to measure the age of the relatively old stellar populations. 
$\chi_{CaII}^2$ is defined as:
\begin{equation}
\chi_{CaII}^2 = (D(CaII)_{obs} - D(CaII)_{template})^2/\sigma_{D(CaII)}^2
\end{equation}  
where $\sigma_{D(CaII)}$ is the observational error.

The luminosity absorbed by dust should mostly be re-emitted in the far-IR. If the energy source of 
the galaxy is dominated by star formation, the absorbed luminosity should match with the far-IR luminosity.
We include this effect as follows:
\begin{equation}
\chi_{fir}^2 = (F_{ir} - F_{absorbed})^2/\sigma_{ir}^2
\end{equation}  
where $F_{ir}$, $F_{absorbed}$, and $\sigma_{ir}$ are the observed flux in 8-1000\micron, the 
absorbed flux in the UV to near-IR, and the observed error of $F_{ir}$, respectively.It is noted that 
$F_{absorbed}$ is calculated for a given $b$ which is determined by $\partial\chi^2_{SED}/\partial b = 0$. 

\begin{figure*}
%%\epsscale{0.8}
%%\includegraphics[angle=0,scale=1.0]{k3_ulirgs.ps}
\includegraphics[width=168mm]{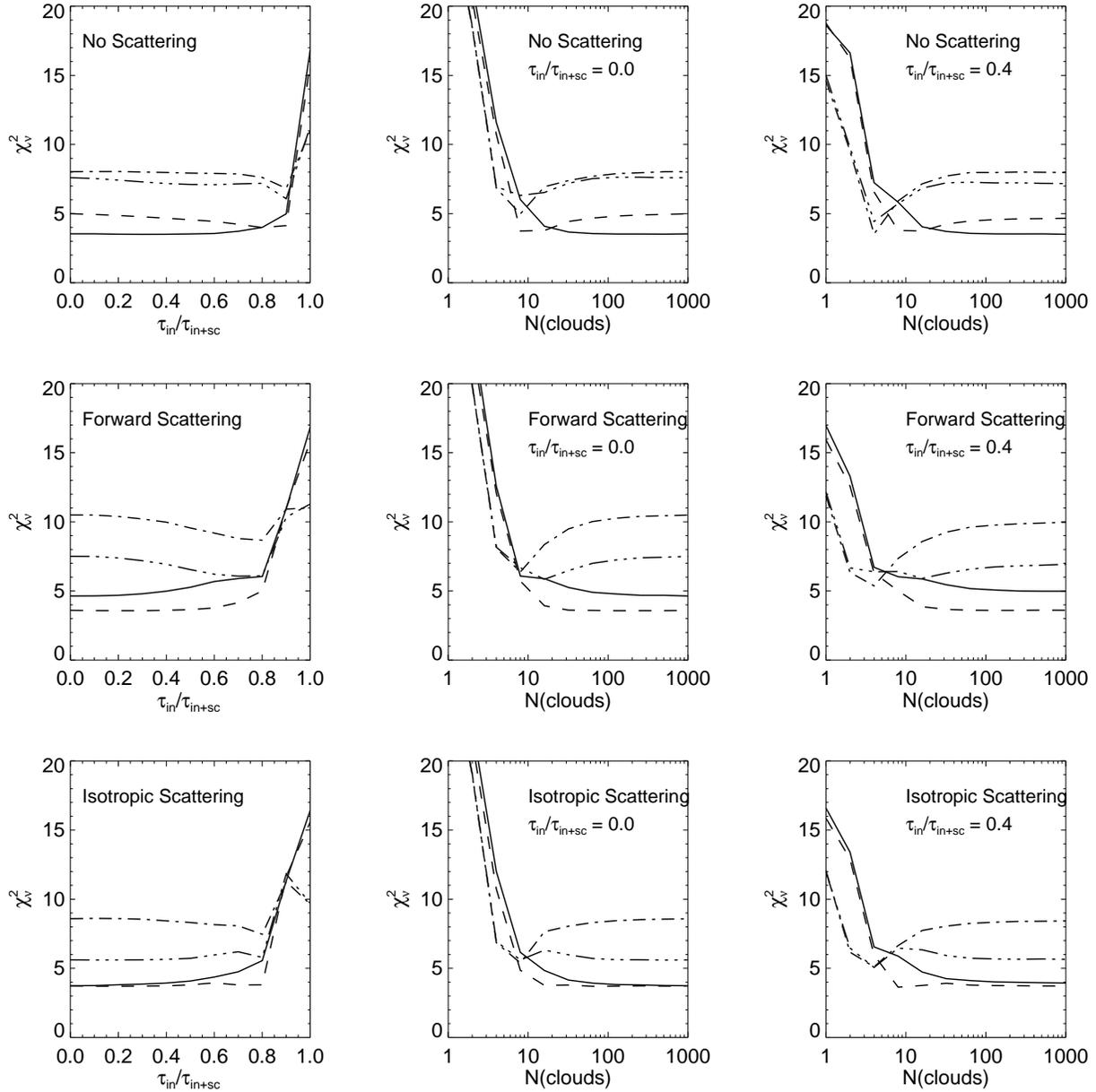}
\caption{The {\it left} column panels show $\chi^2_{\nu}(\nu = 12)$ 
as a function of $\tau_{in}$/($\tau_{in}+\tau_{sc})$
for no scattering ({\it top}), forward scattering ({\it middle}), and 
isotropic scattering ({\it bottom}), where $\tau_{in}$ is the optical depth for 
internal dust and  $\tau_{sc}$ for screen dust. The uniform screen is assumed. 
In case of no scattering, forward scattering is assumed for 
internal dust. {\it Solid} lines represent SN dust with 8 - 30 M$_{\sun}$, 
{\it dash} SN dust with 8 - 40 M$_{\sun}$, {\it dash dot} MW dust, and {\it dash dot dot} pseudo 
MW dust without the 2175 \AA\ feature. 
The {\it middle} column panels are same as the left panels, but as a function 
of the number of clumps in the line of sight contained in the screen for no internal dust 
(i.e., $\tau_{in}$/($\tau_{in}+\tau_{sc}$)  = 0.0).
The {\it right} column panels are same as the left panels, but as a function 
of the number of clumps in the line of sight contained in the screen for 
$\tau_{in}$/($\tau_{in}+\tau_{sc}$) = 0.4. $\chi^2_{\nu} = 5.1$ results if the Calzetti extinction curve 
is applied.
}
\label{f_chi2}
\end{figure*}

\section[]{Results}

First of all, the Calzetti extinction models, which have a uniform foreground screen,  yield 
the best fit-model having the reduced $\chi$-square $\chi^2_{\nu}$ = 5.1 with $\chi^2_{[OII]}$ = 0.0, 
$\chi^2_{CaII}$ = 0.3, and $\chi^2_{fir}$ = 0.9. This model has a parameter set of 
the optical depth $\tau_{ext}(0.3\micron)$ = 5.3, 
the age $t$ = 2.0 $\times$ 10$^8$ yr with the star formation timescale of 
$t_{sfr}$ = 6.3 $\times$ 10$^7$ yr and the metallicity in stellar atmosphere 
$Z = 1 Z_{\sun}$, $EW([OII])$ = 29 \AA, $D(CaII)$ = 0.09, and $F_{absorbed}$ = 
6.6 $\times$ 10$^{-13}$ ergs cm$^{-2}$s$^{-1}$. This is our reference to which the goodness of 
our other models will be judged.

The left column panels of Figure \ref{f_chi2} plot $\chi^2_{\nu}$ as a function of 
$\tau_{in}/(\tau_{in} + \tau_{sc})$ for the uniform foreground screen mixed with internal dust.
$\tau_{in}$ and $\tau_{sc}$ are $\tau_{ext}$ for internal 
dust and screen dust, respectively. For the pure internal dust geometry, 
$\tau_{in}/(\tau_{in} + \tau_{sc})$ = 1, 
while the pure foreground screen geometry, $\tau_{in}/(\tau_{in} + \tau_{sc})$ = 0.
The scattering properties are assumed to be the no-scattering, the isotropic scattering 
($g$ = 0) , or the forward-only scattering ($g$ = 1). In case of no scattering, 
forward scattering is assumed for internal dust.
 
The pseudo MW extinction curve, which is made by removing the 2175 \AA\ feature from the MW extinction 
curve, has a smaller $\chi^2_{\nu}$ than the MW extinction curve, suggesting the 2175 \AA\ feature is 
very weak or unrecognizable in SST J1604+4304.
Because of the large $\chi^2_{\nu}$ values\footnote{
The approximate values of the increment of $\chi_{\nu}^2$ are 1.14, 1.75, and 2.36 for 
68\%(1$\sigma$), 95\%(2$\sigma$), and  99.6\%(3$\sigma$) confidence where the degree of freedom 
is $\nu$ = 12 \citep{avni76}.} relative to the Calzetti and SN extinction curves, 
the MW extinction curve is unlikely regardless of the presence of the 2175\AA\ feature. 
The pure internal dust geometry 
can be ruled out regardless of the choice of extinction curves. Another difficulty of the pure internal 
geometry is that too much dust is required to reproduce the observed SED (i.e., $\tau_{ext}$ = 70 - 200).
The estimated dust mass is far greater than that observed by a factor of 10 or even more. 
The general trend is that 
$\chi^2_{\nu}$ becomes smaller as $\tau_{in}/(\tau_{in} + \tau_{sc})$ approaching to zero, indicating 
that the extinction due to internal dust is a small fraction of the total extinction as already found 
in highly reddened HII regions (i.e.,\citealt{natta84}). The two SN extinction curves have 
$\chi^2_{\nu}$ less than that of the Calzetti extinction curve (i.e., $\chi^2_{\nu}$ = 5.1) 
for a typical range of $\tau_{in}/(\tau_{in} + \tau_{sc})$ = 0.0 - 0.7.     

The middle and right column panels of Figure \ref{f_chi2} plot $\chi^2_{\nu}$ as a function of 
the number of clumps along the line of sight for the clumpy foreground screen; Pure clumpy foreground 
screens (i.e.,$\tau_{in}/(\tau_{in} + \tau_{sc})$ =0) are shown in the middle column panels and 
foreground screens with internal dust having 
$\tau_{in}/(\tau_{in} + \tau_{sc})$ = 0.4 in the right column panels. 
The MW extinction curves with and without the 2175 \AA\ feature have larger $\chi^2_{\nu}$ than that of 
the Calzetti extinction curve except for clumpy screens with the average number of clumps 
of $N$ = 4 (see the upper right panel). 
The overall trends are that small $\chi^2_{\nu}$ values (i.e. $\chi^2_{\nu} < 5.1$) 
are found in larger $N$. Uniform dust screens have been explicitly or implicitly assumed
in many cases in the literature. These figures indicate that uniform dust screens are good approximations 
to the real distribution of dust. The two SN extinction curves have  $\chi^2_{\nu}$ smaller than 
that of the Calzetti extinction curve for large $N$.

\begin{figure}
%%\epsscale{0.8}
\includegraphics[width=84mm]{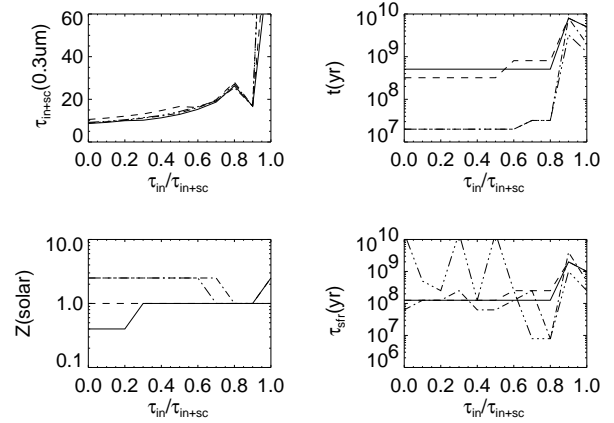}
\caption{Sample illustration of the total optical depth $\tau_{in} + \tau_{sc}$ at 0.3\micron\ 
({\it upper left}), the metallicity $Z$ in units of the solar value ({\it bottom left}), 
the time after the onset of the first star formation $t$(yr) ({\it upper right}), and 
the $e$-folding timescale of the star formation $t_{sfr}$ as a function of the ratio of 
the optical depth of internal dust to internal+screen dust $\tau_{in}/(\tau_{in} + \tau_{sc})$.
The isotropic scattering ($g = 0$)is assumed.{\it Solid} lines represent SN dust with 8 - 30 M$_{\sun}$, 
{\it dash} SN dust with 8 - 40 M$_{\sun}$, {\it dash dot} MW dust, and {\it dash dot dot} pseudo MW dust 
without the 2175 \AA\ feature. 
} 
\label{f_param}
\end{figure}

\begin{figure}
%%\epsscale{0.8}
\includegraphics[width=84mm]{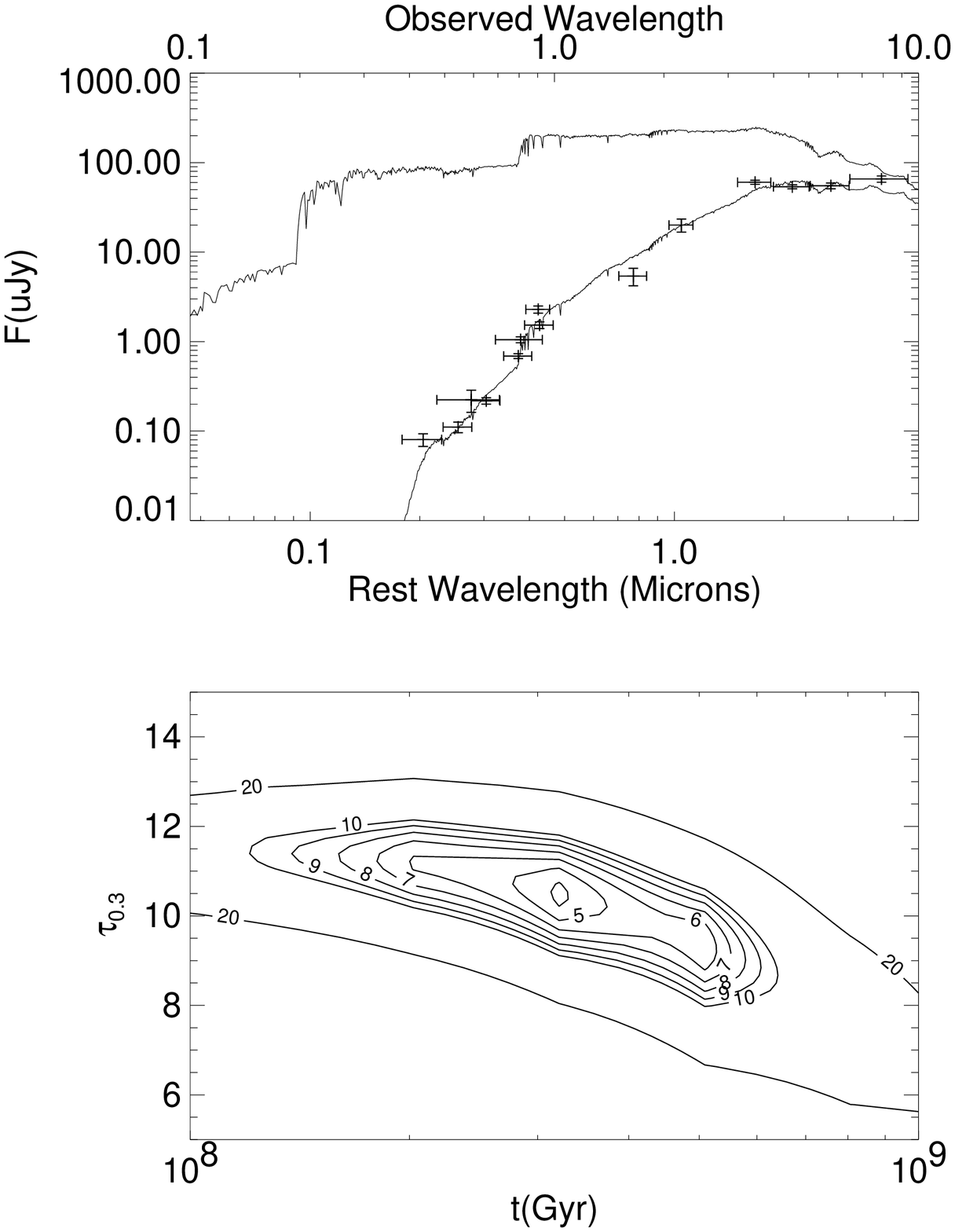}
\caption{Sample illustration of model SEDs. The illustrated model is obtained by assuming
the SN 8-40 M$_{\sun}$ extinction curve with the isotropic scattering without internal dust.
This model has $\chi^2_{\nu}$ = 3.7, $\tau_{sc}$ = 10.2 at 0.3\micron, $t$ = 5.1 $\times 10^8$ yr,
$t_{sfr}$ = 2.5 $\times 10^8$ yr, and $Z = 1 Z_{\sun}$.
The {\it top} panel shows the model spectrum along with the observed SED. The upper curve is the 
unobscured model spectrum.
The {\it bottom} panel shows $\chi^2_{\nu}$ as a function of $t$ and $\tau_{sc}$ at 0.3\micron. 
} 
\label{f_sed}
\end{figure}

Figure \ref{f_param} gives typical features of parameters as a function of 
$\tau_{in}/(\tau_{in} + \tau_{sc})$, taking isotropic scattering models as an example.
Those parameters are the total optical depth
$\tau_{total}(0.3\micron)$(= $\tau_{in}+ \tau_{sc}$ at 0.3\micron), the metallicity $Z$, 
the time after the onset of the first star formation $t$, and 
the $e$-folding timescale $t_{sfr}$. Because 
the effect of scattering reduces the effective optical depth $\tau_{eff}$ as seen in 
equations (7) to (9), scattering dust screens require a larger optical depth than no-scattering 
screens; taking models without internal dust , 
$\tau_{ext}(0.3\micron) \sim 5$ for the no-scattering screen, 
$\tau_{ext}(0.3\micron) \sim 8$ for the isotropic scattering, and 
$\tau_{ext}(0.3\micron) \sim 15$ for the forward-only scattering. 
The optical depth of the SN 8-40 M$_{\sun}$ extinction curve requires larger optical depths than
those of the other curves. This is because albedo is greater in the SN 8-40 M$_{\sun}$ curve than 
the others (see Table \ref{t_kappa}).
$Z$ is smaller on the SN curves than the MW curve. On the other hand, $t$ is older on the SN curves 
than the MW curves. The older $t$ on the SN curves can be attributed to the fact that the slope 
of the SN curves are flatter than that of the MW curve. To reproduce the redness of the SED, 
the steep MW curve requires bluer and younger stellar populations.

Figure \ref{f_sed} shows an example of model SEDs along with the observed SED. 
It is noted again that the longest wavelength data (i.e., IRAC 8.0\micron) is not used for 
$\chi^2_{\nu}$-fitting. The sample model is obtained by assuming
the SN 8-40 M$_{\sun}$ extinction curve with the isotropic scattering without internal dust.
The fitted parameters are $\tau_{sc}(0.3\micron)$ = 10.5, $t$ = 3.2 $\times 10^8$ yr,
$t_{sfr}$ = 1.3 $\times 10^8$ yr, and $Z = 1 Z_{\sun}$.
This model has $\chi^2_{\nu}$ = 3.7 with $\chi^2_{[OII]}$ = 0.4, $\chi^2_{CaII}$ = 0.6, and
$\chi^2_{fir}$ = 2.0, providing a smaller $\chi^2_{\nu}$ than those with the Calzetti extinction 
curve. Comparing with $\chi^2_{\nu}$ = 5.1,  $\chi^2_{[OII]}$ = 0.0, $\chi^2_{CaII}$ = 0.3, and
$\chi^2_{fir}$ = 0.9 for the best-fit model with the Calzetti extinction curve, the reduction in 
$\chi^2_{\nu}$ in this SN model is attributed to the reduction in $\chi^2_{SED}$, indicating that 
the reddening by the SN extinction curve reproduces the SED better than the Calzetti extinction curve.

\section[]{SN II extinction curve}

As shown so far, the SN extinction curves synthesized by \citet{hirashita05} reasonably 
reproduce the observed characteristics of ULIRG SST J1604+4303. The goodness of fit is comparable 
to or better than those with the Calzetti extinction curve. The SN extinction curves must be 
useful to study young dusty galaxies not only at high-redshift but also 
at modest-redshift ($z \sim 1$). We thus present the details of the 
SN extinction curves from the UV to submillimeter.

Figure \ref{f_kappa} shows $\kappa_{abs}(\lambda)$ of the SN dust along with that of the MW dust, 
where $\kappa_{abs}(\lambda)$ is the mass absorption cross section in units of cm$^2$g$^{-1}$. 
Table \ref{t_kappa} gives $\kappa_{ext}(\lambda)$ the mass extinction cross section, 
$\kappa_{abs}(\lambda)$ the mass absorption cross section,
and albedo for the SN dust; SN 8-30 M$_{\sun}$ dust in 2-4 columns and SN 8-40 M$_{\sun}$ dust 
in 5-7 columns . It is noted that the optical constant of crystalline SiO$_2$ is used in the 
models by \citet{hirashita05}. As a result, the SN curves have spiky features
between 9 and 21 \micron. The contribution of crystalline SiO$_2$ to the SN extinction curve is 
a few percent at most in the mid- to far-IR range except the spikes at 8.7, 12.25, and 20.5 \micron. 
We smooth the calculated mass extinction coefficient of Si grains around 1 $\mu$m
to remove small and somehow artificial ripple features arising from large Si grains.

As shown in Figure \ref{f_curves}, the SN extinction curves are flatter than the others:
$R_V$ (= $A_V/E_{B-V}$) is 7.8 for the 8-30 M$_{\sun}$ curve and 5.8 for the 8-40 M$_{\sun}$,
while $R_V$ = 4.05 for the Calzetti curve and $R_V$ = 3.1 for the MW curve. 
In other words, for a given amount of reddening, the extinction is greater in the SN curves 
than the others. 
As suggested in Figure \ref{f_kappa}, the silicate feature around 9.7 \micron\ is much less 
in the SN dust than in the MW dust. This feature 
could be used to distinguish the SN extinction curve from the others by means of mid-IR 
observations in the future.  

Another striking feature of the SN curve is the behavior of $\kappa_{abs}(\lambda)$ from 40 to
1000 \micron. Assuming a simple analytic form of 
$\kappa_{abs}(\lambda) \propto \nu^{\beta}$, we have $\beta$ $\simeq 1.6$ for the SN dust, 
while $\beta$ $\simeq 2 $ for the MW dust. Single-temperature dust spectral 
distributions of the form $\nu^{\beta} B_{\nu}(T_{dust})$, are widely used to fit observed SED 
in the far-IR and submillimeter, where $B_{\nu}(T_{dust})$ is the Planck function. 
However, fitting only a few datapoints results in a significant correlation between $\beta$ 
and $T_{dust}$ that can account for the data \citep{blain03}:
higher $T_{dust}$ with smaller $\beta$; $T_{dust}$ derived with the SN dust is 13\% higher than 
that with the MW dust. Assuming the SN dust, we have $T_{dust} = 38.3 \pm 1.4 K$ for SST J1604+4304 
which is a common characteristics of 104 $IRAS$ bright 
galaxies observed at 850 \micron\ by \citet{dunne00}, where
the best fit was found for $T_{dust} = 35.6 \pm 4.9$ K and $\beta = 1.3 \pm 0.2$. 

\section[]{Dust in SST J1604+4304}

Using $T_{dust} = 38.3 \pm 1.4 K$ combined with the observed 160\micron\ flux density and 
$\kappa_{abs}(80\micron) = 22 $cm$^2$ g$^{-1}$ in Table \ref{t_kappa}, implies that the dust mass 
$M_{dust}$ in SST J1604+4303 is $2.2 \pm 1.1 \times 10^8$ M$_{\sun}$. Consider a spherical shell 
with a radius of $r$ and a thickness of $\delta r$, and suppose that internal dust fills  
inside the sphere with $r$ and screen dust in the shell. The mass of the internal dust can be simply 
approximated to $M_{in}$ = $4\pi(\tau_{in}/\kappa_{ext})r^2/3$ and that of the screen dust to 
$M_{sc}$ = $4\pi(\tau_{sc}/\kappa_{ext})r^2$, thus $M_{in}/M_{sc} = (\tau_{in}/\tau_{sc})/3$. 
The best model for the isotropic scattering uniform screen without internal dust has 
$\tau_{ext}(0.3\micron)$ = 10.5. In this case, the inferred radius of the spherical shell is 
about 3.3 kpc or 0.4 arcsec which is comparable to the optical size of SST J1604+4303.

If the star formation history of SST J1604+4303 is described by $t$ = 3.2 $\times 10^8$ yr,
$t_{sfr}$ = 1.3 $\times 10^8$ yr, and $Z = 1Z_{\sun}$, 
then the luminosity to mass ratio is 7.6, implying a stellar 
mass of 2.4 $\times 10^{11}$ M$_{\sun}$. During the last 3.2 $\times 10^8$ yr, $\sim 2.0 \times 10^9$ 
SNe II exploded in this galaxy. If the dust was mainly supplied by SN explosions, 
then we have a dust production rate of $ \sim 0.1$ M$_{\sun}$/SN by simply dividing the dust mass 
by the number of SNe II. 

Although $\chi^2_{\nu}$  is reduced by adopting the synthesized SN extinction 
curves, the goodness of fit is far from the perfect fitting(i.e.,$\chi^2_{\nu} \sim 1$). 
This may come from various reasons. 
For example, the models for dust formation in SNe II are still coarse and the contribution from 
the dust formed in AGB stars needs to be included, or the stellar 
population models are too simple and the real stellar population may be more complex and 
consist of stars with different ages with different metallicities. 

{\citet{nozawa07} take into account the effect of reverse shock
destruction and \citet{hirashita08} calculate the extinction curve based on their
result. The extinction curves after reverse shock destruction are generally flatter than
those of \citet{hirashita05}, since small grains are selectively destroyed. In this
paper, we simply adopt the extinction curves by \citet{hirashita05}. In other words, the effect of
reverse shock destruction is not taken into account. Indeed, the flat extinction
curves after reverse shock destruction are not consistent with the large reddening
required for SST J1604+4304. This implies that the reverse shock destruction
may not be efficient or that the small
grain production by, for example, shattering in warm ISM is
occurring \citep{hirashita10}, in which case the extinction curve after shattering
is similar to that used in this paper. Moreover, the extinction curve by
\citet{hirashita05} is supported by
observations in the sense that it is consistent with
the extinction curve derived by \citet{maiolino04} for a high-$z$ BAL QSO
at $z=6.2$.
Note that the \citet{maiolino04}'s extinction curve is also consistent with
a high-$z$ ($z=6.3$) gamma-ray burst \citep{stratta07}.
  
We have shown that a flatter extinction curve (or
a larger $R_V$) than the MW curve fits the SED of SST
J1604+4304 better. A flat extinction curve may also be
realized if dust growth occurs in dense clouds \citep{draine03}. 
Thus, it is worth investigating whether or not a
flat extinction curve predicted from the grain growth
fits the observed SED of SST J1604+4304. This is left
for future work.

\begin{table*}
 \centering
%% \begin{minipage}{70mm}
  \caption{Extinction cross sections of the SN II dust models}
  \begin{tabular}{cccccccc}
  \hline
       & \multicolumn{3}{c}{SN 8-30 M$_{\sun}$} & & \multicolumn{3}{c}{SN 8-40 M$_{\sun}$} \\
\cline{2-4} \cline{6-8} \\
 $\lambda(\micron)$ & $\kappa_{ext}$(cm$^2$ g$^{-1}$) & $\kappa_{abs}$(cm$^2$ g$^{-1}$) & albedo & 
                    & $\kappa_{ext}$(cm$^2$ g$^{-1}$) & $\kappa_{abs}$(cm$^2$ g$^{-1}$) & albedo \\ 
  \hline
    %% created by k3_ table_paper.dat
%% 1st column --- wavelength (microns)
%% 2nd column --- mass extinction coefficients in cm^2/g for SN 8-30 Msun
%% 3rd column --- mass absorption coefficients in cm^2/g for SN 8-30 Msun
%% 4th column --- albedo for SN 8-30 Msun
%% 5th column --- mass extinction coefficients in cm^2/g for SN 8-40 Msun
%% 6th column --- mass absorption coefficients in cm^2/g for SN 8-40 Msun
%% 7th column --- albedo for SN 8-40 Msun
%%
 6.67e-02 &    4.69e+04 &    2.34e+04 &    5.01e-01& &   4.86e+04 &    2.37e+04 &    5.11e-01\\
 9.12e-02 &    4.73e+04 &    2.34e+04 &    5.05e-01& &   4.95e+04 &    2.44e+04 &    5.07e-01\\
 1.00e-01 &    4.71e+04 &    2.30e+04 &    5.12e-01& &   4.95e+04 &    2.44e+04 &    5.07e-01\\
 1.20e-01 &    5.07e+04 &    2.57e+04 &    4.93e-01& &   5.23e+04 &    2.64e+04 &    4.96e-01\\
 1.40e-01 &    5.09e+04 &    2.43e+04 &    5.23e-01& &   5.25e+04 &    2.47e+04 &    5.29e-01\\
 1.60e-01 &    4.68e+04 &    2.03e+04 &    5.67e-01& &   4.97e+04 &    2.12e+04 &    5.74e-01\\
 1.80e-01 &    4.04e+04 &    1.67e+04 &    5.87e-01& &   4.54e+04 &    1.61e+04 &    6.46e-01\\
 1.90e-01 &    3.87e+04 &    1.57e+04 &    5.94e-01& &   4.29e+04 &    1.40e+04 &    6.74e-01\\
 2.00e-01 &    3.79e+04 &    1.50e+04 &    6.04e-01& &   4.14e+04 &    1.23e+04 &    7.02e-01\\
 2.10e-01 &    3.77e+04 &    1.46e+04 &    6.12e-01& &   4.04e+04 &    1.13e+04 &    7.19e-01\\
 2.20e-01 &    3.77e+04 &    1.45e+04 &    6.15e-01& &   3.96e+04 &    1.08e+04 &    7.26e-01\\
 2.30e-01 &    3.77e+04 &    1.46e+04 &    6.14e-01& &   3.88e+04 &    1.07e+04 &    7.24e-01\\
 2.50e-01 &    3.75e+04 &    1.44e+04 &    6.16e-01& &   3.72e+04 &    1.04e+04 &    7.21e-01\\
 3.00e-01 &    3.40e+04 &    1.39e+04 &    5.92e-01& &   3.10e+04 &    9.80e+03 &    6.84e-01\\
 3.55e-01 &    3.17e+04 &    1.32e+04 &    5.82e-01& &   2.73e+04 &    9.00e+03 &    6.70e-01\\
 3.64e-01 &    3.14e+04 &    1.31e+04 &    5.81e-01& &   2.68e+04 &    8.89e+03 &    6.68e-01\\
 4.41e-01 &    2.80e+04 &    1.20e+04 &    5.71e-01& &   2.26e+04 &    8.15e+03 &    6.40e-01\\
 4.69e-01 &    2.69e+04 &    1.16e+04 &    5.69e-01& &   2.14e+04 &    7.84e+03 &    6.34e-01\\
 5.47e-01 &    2.48e+04 &    1.08e+04 &    5.65e-01& &   1.93e+04 &    7.30e+03 &    6.22e-01\\
 6.17e-01 &    2.21e+04 &    9.88e+03 &    5.53e-01& &   1.70e+04 &    6.64e+03 &    6.10e-01\\
 6.42e-01 &    2.10e+04 &    9.66e+03 &    5.40e-01& &   1.61e+04 &    6.51e+03 &    5.96e-01\\
 6.49e-01 &    2.08e+04 &    9.60e+03 &    5.39e-01& &   1.60e+04 &    6.47e+03 &    5.95e-01\\
 7.00e-01 &    2.03e+04 &    9.21e+03 &    5.46e-01& &   1.55e+04 &    6.19e+03 &    6.00e-01\\
 7.48e-01 &    1.95e+04 &    8.90e+03 &    5.45e-01& &   1.49e+04 &    5.98e+03 &    5.99e-01\\
 8.02e-01 &    1.86e+04 &    8.55e+03 &    5.40e-01& &   1.41e+04 &    5.73e+03 &    5.94e-01\\
 8.66e-01 &    1.75e+04 &    8.16e+03 &    5.34e-01& &   1.32e+04 &    5.46e+03 &    5.86e-01\\
 8.93e-01 &    1.70e+04 &    7.99e+03 &    5.29e-01& &   1.27e+04 &    5.36e+03 &    5.79e-01\\
 9.00e-01 &    1.68e+04 &    7.95e+03 &    5.27e-01& &   1.26e+04 &    5.33e+03 &    5.77e-01\\
 1.00e+00 &    1.51e+04 &    7.36e+03 &    5.12e-01& &   1.13e+04 &    4.95e+03 &    5.62e-01\\
 1.22e+00 &    1.26e+04 &    6.08e+03 &    5.19e-01& &   9.64e+03 &    4.12e+03 &    5.72e-01\\
 1.63e+00 &    9.48e+03 &    4.48e+03 &    5.27e-01& &   7.52e+03 &    3.03e+03 &    5.96e-01\\
 2.19e+00 &    5.83e+03 &    3.07e+03 &    4.74e-01& &   4.82e+03 &    2.05e+03 &    5.75e-01\\
 2.75e+00 &    4.14e+03 &    2.42e+03 &    4.14e-01& &   3.49e+03 &    1.62e+03 &    5.35e-01\\
 3.45e+00 &    2.73e+03 &    1.87e+03 &    3.15e-01& &   2.40e+03 &    1.17e+03 &    5.11e-01\\
 3.80e+00 &    2.30e+03 &    1.59e+03 &    3.06e-01& &   2.02e+03 &    9.94e+02 &    5.08e-01\\
 4.75e+00 &    1.66e+03 &    1.13e+03 &    3.21e-01& &   1.53e+03 &    6.87e+02 &    5.51e-01\\
 6.00e+00 &    9.46e+02 &    6.35e+02 &    3.29e-01& &   7.70e+02 &    3.72e+02 &    5.17e-01\\
 7.00e+00 &    8.27e+02 &    5.46e+02 &    3.40e-01& &   7.37e+02 &    3.25e+02 &    5.58e-01\\
 8.00e+00 &    6.33e+02 &    4.72e+02 &    2.53e-01& &   4.80e+02 &    3.10e+02 &    3.53e-01\\
 8.28e+00 &    6.21e+02 &    4.73e+02 &    2.39e-01& &   4.70e+02 &    3.33e+02 &    2.94e-01\\
 8.71e+00 &    1.56e+03 &    1.25e+03 &    1.97e-01& &   1.61e+03 &    1.38e+03 &    1.46e-01\\
 9.17e+00 &    6.34e+02 &    4.99e+02 &    2.13e-01& &   5.56e+02 &    4.62e+02 &    1.69e-01\\
 9.70e+00 &    6.63e+02 &    5.69e+02 &    1.42e-01& &   6.42e+02 &    5.71e+02 &    1.10e-01\\
 1.13e+01 &    4.69e+02 &    3.61e+02 &    2.29e-01& &   3.86e+02 &    3.22e+02 &    1.64e-01\\
 1.20e+01 &    3.93e+02 &    3.25e+02 &    1.74e-01& &   3.20e+02 &    2.79e+02 &    1.27e-01\\
 1.38e+01 &    2.56e+02 &    2.29e+02 &    1.02e-01& &   1.94e+02 &    1.79e+02 &    8.14e-02\\
 1.53e+01 &    3.22e+02 &    2.97e+02 &    7.65e-02& &   3.49e+02 &    3.38e+02 &    3.37e-02\\
 1.61e+01 &    6.18e+02 &    6.01e+02 &    2.80e-02& &   8.32e+02 &    8.23e+02 &    1.14e-02\\
 1.70e+01 &    5.95e+02 &    5.84e+02 &    1.74e-02& &   8.06e+02 &    7.99e+02 &    8.84e-03\\
 1.79e+01 &    2.74e+02 &    2.67e+02 &    2.68e-02& &   3.01e+02 &    2.96e+02 &    1.61e-02\\
 1.98e+01 &    2.50e+02 &    2.45e+02 &    2.00e-02& &   2.70e+02 &    2.67e+02 &    1.17e-02\\
 2.08e+01 &    2.90e+02 &    2.77e+02 &    4.49e-02& &   3.08e+02 &    3.04e+02 &    1.54e-02\\
 2.19e+01 &    2.03e+02 &    1.99e+02 &    1.97e-02& &   2.12e+02 &    2.10e+02 &    1.07e-02\\
 2.50e+01 &    1.47e+02 &    1.45e+02 &    1.41e-02& &   1.44e+02 &    1.43e+02 &    8.60e-03\\
 4.00e+01 &    5.69e+01 &    5.66e+01 &    4.89e-03& &   4.55e+01 &    4.53e+01 &    3.96e-03\\
 6.00e+01 &    3.30e+01 &    3.30e+01 &    1.61e-03& &   2.68e+01 &    2.68e+01 &    1.34e-03\\
 1.00e+02 &    1.59e+01 &    1.59e+01 &    4.08e-04& &   1.28e+01 &    1.28e+01 &    3.41e-04\\
 1.40e+02 &    1.09e+01 &    1.09e+01 &    1.96e-04& &   8.45e+00 &    8.45e+00 &    1.47e-04\\
 2.00e+02 &    5.08e+00 &    5.08e+00 &    8.02e-05& &   4.08e+00 &    4.08e+00 &    6.63e-05\\
 3.50e+02 &    2.03e+00 &    2.03e+00 &    2.25e-05& &   1.61e+00 &    1.61e+00 &    1.87e-05\\
 8.50e+02 &    4.38e-01 &    4.38e-01 &    3.09e-06& &   3.57e-01 &    3.57e-01 &    2.42e-06\\
 1.00e+03 &    3.17e-01 &    3.17e-01 &    2.00e-06& &   2.65e-01 &    2.65e-01 &    1.61e-06\\

\hline
\end{tabular}
\medskip \\
 \label{t_kappa}
%%\end{minipage}
\end{table*}

\begin{figure}
%%\epsscale{0.8}
\includegraphics[width=84mm]{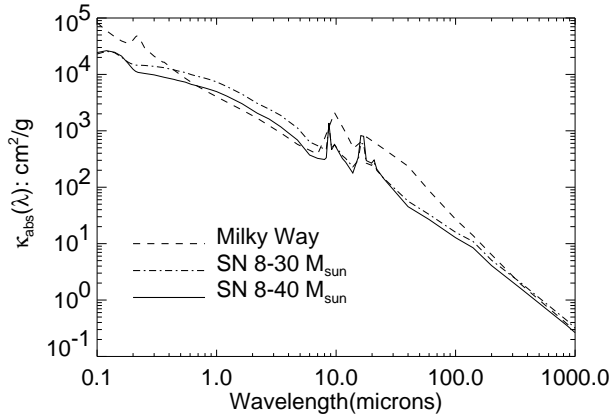}
%%\includegraphics[angle=0,scale=1.0]{k3_closeup.ps}
%%\plotone{c1604_image_zoom.ps}
\caption{Plot of $\kappa_{abs}(\lambda)$ the absorption cross section per dust mass in units of 
cm$^2$/g. Note that spiky features between 9 and 21 \micron\ on the SN curve are mostly attributed to 
crystal SiO$_2$ and MgO; spiky features of crystal SiO$_2$ occur at 8.7, 12.25, \& 20.5 \micron\ while
a spike MgO feature at 16.5 \micron} 
\label{f_kappa}
\end{figure}

\section[]{Conclusions}

We have applied the SN extinction curves to reproduce the observed properties of SST J1604+4304 
which is a young IR galaxy at $z \sim 1$. The SN extinction curves used in this work was obtained 
from models of unmixed ejecta of SNe II for the Salpeter IMF with a mass range from 8 to 30 $M_{\sun}$ or
8 to 40 $M_{\sun}$. These curves are compared with the MW and Calzetti extinction curves.
The data used in the $\chi^2$ fitting methods 
are the fluxes in 13 photometric bands from $B$ to IRAC 5.8\micron, the equivalent width of the 
[OII]$\lambda$ 3727, the D(CaII) index, and the IR flux in 8-1000\micron. 

The effect of dust distributions on the attenuation of starlight was investigated by performing 
the $\chi^2$ fitting method with various dust distributions. These are the commonly used uniform dust 
screen, the clumpy dust screen, and the internal dust geometry. In addition to these geometries, 
we have investigated three scattering properties, namely, no-scattering, isotropic scattering, 
and forward-only scattering. Judging from the  $\chi^2$ values, 
the uniform screen models with any scattering property provide good approximations to the real 
dust geometry. Internal dust is inefficient to attenuate starlight and thus cannot be the dominant 
source of the extinction.

We have found that the SN extinction curves reproduce the data of SST J1604+4304 comparable or 
better than the Calzetti extinction curve. The MW extinction curve is too steep and is not 
satisfactory agreement with the data unless several dusty clumps are in the line of sight. 
This trend may be explained by 
the abundance of SN-origin dust in these galaxies; SN dust is the most abundant in the 
young IR galaxy at $z \sim 1$, abundant in local starbursts, and less abundant in the Galaxy.
If dust in SST J1604+4304 is dominated by SN dust, the dust production rate is 
$\sim 0.1$ M$_{\sun}$/SN.

We provide the extinction cross sections of the SN dust along with albedo from UV to 
submillimeter wavelengths in Table \ref{t_kappa} and on-line material table.
 
\section*{Acknowledgments}

We thank the anonymous referee for very helpful and constructive comments.
This work has been supported in part by Grants-in-Aid for Specially Promoted Research on 
Innovative Areas(22111503), Specially Promoted Research (20001003), 
and the Promotion of Science (18104003, 20340038, 22684004) from JSPS.
H.H. is supported by NSC grant 99-2112-M-001-006-MY3.

\appendix
\section{Two stream approximation for radiation transfer in a dusty slab}

The two stream approximation in radiation transfer was applied to a spherically 
symmetric dust envelope surrounding a star by \citet{code73}. Here we consider a
plane-parallel dust slab.

We measure distance $z$ outward for a plane-parallel slab, then

\begin{eqnarray}
\frac{\mu}{\rho} \frac{dI_{\nu}(z,\mu)}{dz}= j_{\nu} - \kappa_{\nu}I_{\nu}(z,\mu)+
\frac{\sigma_{\nu}}{4 \pi}
\int_{4\pi} p(\mu,\mu')I_{\nu}(z,\mu) d\omega'
\end{eqnarray}
where  $I_{\nu}(z,\mu)$ is the specific intensity, $\mu=\cos\theta$ 
the directional cosine, $\nu$ the frequency,
$\kappa_{\nu}$ the mass extinction coefficient, $\sigma_{\nu}$
the mass scattering coefficient, $\rho$ the mass density of dust, $j_{\nu}$ 
the ratio of the stellar emission per unit dust mass. 
The phase function  $p(\mu, \mu')$ is defined by 

\begin{eqnarray}
\int_{4\pi} p(\mu,\mu')d\omega' = 4\pi  \hspace{0.5cm}
\textrm{or} \hspace{0.4cm} \frac{1}{2}\int_{-1}^{1}p(\mu,\mu')d\mu'=1
\end{eqnarray}

By introducing the optical depth $\tau_{\nu}$ which is defined by 
$d\tau_{\nu}= 
-\rho\kappa_{\nu} dz$, the radiation transfer equation is reduced to 

\begin{eqnarray}
\mu \frac{d I_{\nu}(\tau,\nu)}{d\tau_{\nu}}=  
-j_{\nu}/\kappa_{\nu} + I_{\nu} - \frac{\gamma_{\nu}}{2}
\int_{-1}^{1}p(\mu,\mu')I_{\nu}(\tau,\mu')d\mu'
\end{eqnarray}

\noindent
where $\gamma_{\nu}=\sigma_{\nu}/\kappa_{\nu}$ is the albedo. In what
follows, we omit the subscript $\nu$ for simplicity. 

\subsection{Slab illuminated at the back side}

Consider a slab which contains no emitters and illuminated at the back side, 
i.e., $j_{\nu}$ = 0. In the two stream
approximation, the specific intensity $I(\tau,\mu)$ is represented by
the specific intensities at two specified directions at $\pm\mu$; 
$I^+(\tau)=I(\tau,\mu)$ and $I^{-}(\tau)=I({\tau, -\mu})$ for
$0\le\mu\le1$. $I^+(\tau)$
and $I^{-}(\tau)$ are assumed to be independent of $\mu$ over the upward
and the downward hemispheres, respectively.

According to \citet{code73}, we introduce the forward and backward
scattering coefficients
$\sigma^+$ and $\sigma^{-}$ which are defined by: 

\begin{eqnarray}
\frac{\sigma^+}{\sigma}=\frac{1}{2}\int_0^1p(\mu,\mu')d\mu' 
\hspace{0.5cm}\textrm{and} 
\hspace{0.5cm}\frac{\sigma^{-}}{\sigma}=\frac{1}{2}
\int_0^1p(\mu,-\mu')d\mu' 
\end{eqnarray}

\noindent
respectively. Note that $\sigma^++\sigma^{-}=\sigma$. A modified 
asymmetry factor $g$ is defined by 

\begin{eqnarray}
g=\frac{\sigma^+-\sigma^{-}}{\sigma}
\end{eqnarray}
\noindent
note that $g=1$ for forward scattering, $g=-1$ for backward scattering, 
and $g=0$ for isotropic scattering.
Then using the relations 
 $\sigma^+/\sigma=(1+g)/2$ and $\sigma^{-}/\sigma=(1-g)/2$,  the equations
 (A3) is reduced to: 

\begin{eqnarray}
\mu\frac{dI^+(\tau)}{d\tau}=I^+(\tau)-\gamma\frac{(1+g)}{2}I^+(\tau) 
-\gamma\frac{(1-g)}{2}I^{-}(\tau)
\end{eqnarray}

\begin{eqnarray}
-\mu\frac{dI^{-}(\tau)}{d\tau}=I^{-}(\tau)-\gamma\frac{(1+g)}{2}I^{-}(\tau)
-\gamma\frac{(1-g)}{2}I^+(\tau)
\end{eqnarray}

Here we replace the optical depth along the $z$-axis with the
optical depth along the line of sight $d\tau_{ext} =d\tau/\mu$ 
and introduce new 
variables $x=I^+-I^{-}$ and $y=I^++I^{-}$, the above
equations become simply: 

\begin{eqnarray}
\frac{d x}{d\tau_{ext}} = \left(1-\gamma\right)y \hspace{0.5cm}
\textrm{and} \hspace{0.4cm}
\frac{d y}{d\tau_{ext}} = \left(1-\gamma g \right)x 
\end{eqnarray}
\noindent

Then the solutions of $I^+(\tau_{ext})$ and $I^{-}(\tau_{ext})$ are given by
{\footnotesize
\begin{eqnarray}
I^+(\tau_{ext})=
\frac{1}{2}\left[(\beta+1)A\exp\left(\alpha\tau_{ext}\right)-
(\beta-1)B\exp\left(-\alpha\tau_{ext}\right)\right]
\end{eqnarray} 
}
{\footnotesize
\begin{eqnarray}
I^{-}(\tau_{ext})=
\frac{1}{2}\left[(\beta-1)A\exp\left(\alpha\tau_{ext}\right)-
(\beta+1)B\exp\left(-\alpha\tau_{ext}\right)\right]
\end{eqnarray}
}
\noindent 
where $\alpha^2=(1-\gamma)(1-\gamma g)$ and $\beta=
[(1-\gamma g)/(1-\gamma)]^{1/2}$. Using the boundary conditions of
$I^{-}(0)=0$ and $I^+(\tau_{1ext})=I_0$ 
for determing the constants $A$ and $B$, where 
$\tau_{1ext}$ is the optical depth of the slab along the line of sight,
then the emergent intensity $I^+(0)$ is given by 

\begin{eqnarray}
I^+(0)= \frac{2\beta I_0}{\left(\beta+1\right)\exp\left(\alpha\tau_{1ext}\right)
+\left(\beta-1\right)\exp\left(-\alpha\tau_{1ext}\right)}
\end{eqnarray}

\subsection{Slab containing emitters}

We now consider a slab which contains emitters ,i.e., stars. The slab is 
illuminated by these internal stars only. Then 
the radiation transfer equations for $I^{\pm}$ are simply given by 

\begin{eqnarray}
\mu\frac{dI^+(\tau)}{d\tau} &= I^+(\tau) 
 -\gamma\frac{(1+g)}{2}I^+(\tau) \nonumber \\
&-\gamma\frac{(1-g)}{2}I^{-}(\tau)-\Lambda
\end{eqnarray}

\begin{eqnarray}
-\mu\frac{dI^{-}(\tau)}{d\tau} &=I^{-}(\tau)
-\gamma\frac{(1+g)}{2}I^{-}(\tau) \nonumber \\
&-\gamma\frac{(1-g)}{2}I^+(\tau)-\Lambda
\end{eqnarray}

\noindent
where $\Lambda=\j_{\nu}/\kappa_{\nu}$.  
Then $x=I^+-I^{-}$ and $y=I^++I^{-}$ can be 
determined by solving the following simultaneous equations; 

\begin{eqnarray}
\frac{dx}{d\tau_{ext}}=\left(1-\gamma\right)y-2\Lambda \hspace{0.5cm}
\textrm{and}\hspace{0.5cm} \frac{dy}{d\tau_{ext}}=\left(1-\gamma g\right)x.
\end{eqnarray}

\noindent
The intensities $I^{\pm}(\tau_{ext})$ are given by 

\begin{eqnarray}
I^+(\tau_{ext}) &=
\frac{1}{2}(\beta+1)\left[A-E_0(\tau_{ext})\right]\exp\left(\alpha\tau_{ext}\right) 
\nonumber \\
&-\frac{1}{2}(\beta-1)\left[B-E_1(\tau_{ext})\right]\exp\left(-\alpha\tau_{ext}\right)
\end{eqnarray}

\begin{eqnarray}
I^{-}(\tau_{ext}) &=
\frac{1}{2}(\beta-1)\left[A-E_0(\tau_{ext})\right]\exp\left(\alpha\tau_{ext}\right)
\nonumber \\
&-\frac{1}{2}(\beta+1)\left[B-E_1(\tau_{ext})\right]\exp\left(-\alpha\tau_{ext}\right)
\end{eqnarray}

\noindent
where
\begin{eqnarray}
E_0(\tau_{ext})& =
\int_0^{\tau_{ext}}\Lambda(\tau_{ext}')\exp\left(-\alpha\tau_{ext}'\right) d\tau_{ext}' 
\hspace{0.2cm}\textrm{and}\hspace{0.2cm} \nonumber \\
E_1(\tau_{ext})&=
\int_0^{\tau_{ext}}\Lambda(\tau_{ext}')\exp\left(\alpha\tau_{ext}'\right) d\tau_{ext}'
\end{eqnarray}

Given the boundary conditions of $I^{-}(0)=I^+(\tau_{1ext})=0$,
the emergent intensity $I^+(0)$ is given by:
 {\footnotesize
\begin{eqnarray}
I^+(0)=2\beta
\frac{(\beta+1)E_0(\tau_{1ext})\exp\left(\alpha\tau_{1ext}\right)-
(\beta-1)E_1(\tau_{1ext})\exp\left(-\alpha\tau_{1ext}\right)}
{(\beta+1)^2\exp\left(\alpha\tau_{1ext}\right)-
(\beta-1)^2\exp\left(-\alpha\tau_{1ext}\right)} 
\end{eqnarray}
}

If $\Lambda$ = const against $\tau_{ext}$, 
$I^+(0) \rightarrow  \Lambda\tau_{1eff}$ for $\tau_{1ext} \rightarrow 0$. 
Thus, 
the apparant extinction $\epsilon_{in}$ as defined in equation (1) is:
 {\footnotesize
\begin{eqnarray}
\epsilon_{in}=\frac{2\beta}{\alpha\tau_{1ext}}
\frac{(\beta+1)[\exp\left(\alpha\tau_{1ext}\right)-1]-
(\beta-1)[1-\exp\left(-\alpha\tau_{1ext}\right)]}
{(\beta+1)^2\exp\left(\alpha\tau_{1ext}\right)-
(\beta-1)^2\exp\left(-\alpha\tau_{1ext}\right)} 
\end{eqnarray}
}

Figure \ref{f_appendix} compares equation (A19) with the approximations by 
$\epsilon_{in}$ = $(1-e^{-\tau_{eff}})/\tau_{eff}$ for 
$\tau_{eff} = (1-\gamma)\tau_{ext}$,
$\tau_{eff} = (1-\gamma)^{0.5}\tau_{ext}$, 
and $\tau_{eff} = (1-\gamma)^{0.75}\tau_{ext}$. 
The geometry is internal dust with
$\gamma$ = 0.65 and $g$ = 0. 
As shown in Figure \ref{f_appendix},
$\tau_{eff} = (1-\gamma)^{0.75}\tau_{ext}$ 
approximates to equation (A19) better than 
$\tau_{eff} = (1-\gamma)^{0.5}\tau_{ext}$ for the isotropic scattering.

\begin{figure}
%%\epsscale{0.8}
\includegraphics[width=84mm]{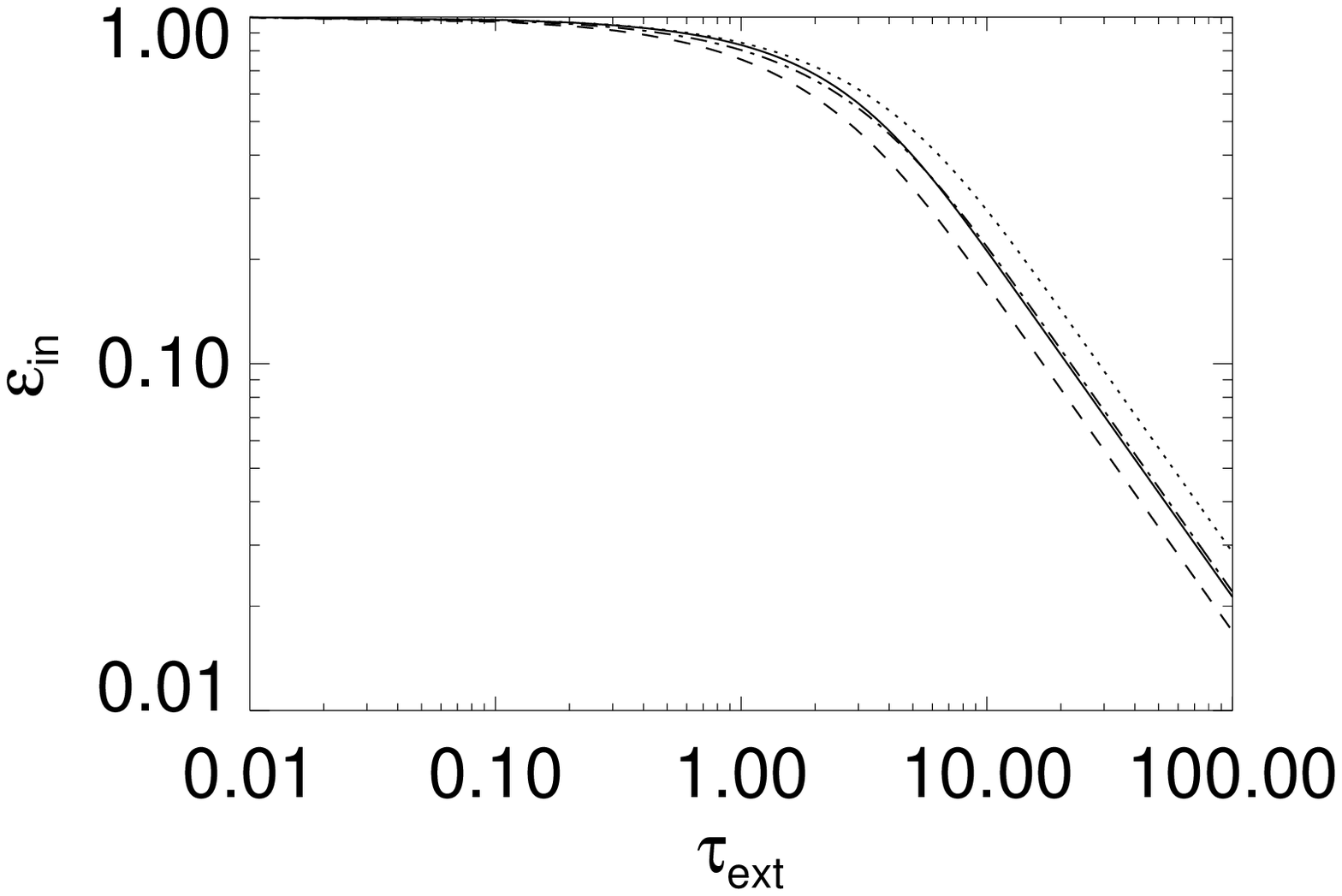}
%%\includegraphics[angle=0,scale=1.0]{k3_closeup.ps}
%%\plotone{c1604_image_zoom.ps}
\caption{Plot of the apparent extinction $\epsilon_{in}$ 
as a function of the optical 
depth along the line of sight $\tau_{ext}$. The geometry is the internal dust with 
$\gamma$ = 0.65 and $g$ = 0 (spherical scattering).
The {\it solid} line represents the solution of equation (A19) for 
$\Lambda$ = const against $\tau_{ext}$. 
The other lines represent the absorption approximated by $\epsilon_{in} $ = 
$(1-e^{-\tau_{eff}})/\tau_{eff}$; the {\it dotted} line for 
$\tau_{eff} = (1-\gamma)\tau_{ext}$,the {\it dashed} line for 
$\tau_{eff} = (1-\gamma)^{0.5}\tau_{ext}$, and the {\it dash dot} line for 
$\tau_{eff} = (1-\gamma)^{0.75}\tau_{ext}$. 
} 
\label{f_appendix}
\end{figure}

\end{document}